\documentclass[fleqn,usenatbib]{mnras}

\usepackage{newtxtext,newtxmath}

\usepackage[T1]{fontenc}
\usepackage{ae,aecompl}

\usepackage{graphicx}	
\usepackage{amsmath}	
\usepackage{amssymb}	
\usepackage{threeparttable}

\title[Li-rich giant stars under scrutiny]{Li-rich giant stars under scrutiny: Binarity, magnetic activity and the evolutionary status after Gaia DR2}

\author[B. F. O. Gon\c{c}alves et al.]{B. F. O. Gon\c{c}alves,$^{1}$\thanks{E-mail: odlavson@fisica.ufrn.br (UFRN)}
J. S. da Costa,$^{2}$
L. de Almeida,$^{1}$
M. Castro$^{1}$ and 
\newauthor J.-D. do Nascimento, Jr.$^{1,3}$
\\
$^{1}$ Departamento de F\'isica, DFTE, Universidade Federal do Rio Grande do Norte, UFRN, 59072-970, Natal, RN, Brazil\\
$^{2}$ Escola de Ci\^encias e Tecnologia, ECT, Universidade Federal do Rio Grande do Norte, UFRN, 59078-970, Natal, RN, Brazil\\
$^{3}$ Harvard-Smithsonian Center for Astrophysics, 60 Garden St., Cambridge, MA 02138, USA
}

\date{Accepted 2020 August 6. Received 2020 August 1; in original form 2019 June 22}

\pubyear{2020}

\begin{document}
\label{firstpage}
\pagerange{\pageref{firstpage}--\pageref{lastpage}}
\maketitle

\begin{abstract}
We present a study of the evolutionary state of a few lithium-rich giant stars based on the Gaia DR2 parallaxes and photometry. We also investigate the chromospheric activity, the presence of a surface magnetic field, and the radial velocity for our sample stars. We analysed both archive and new data. We gathered archive spectra from several instruments, mainly ELODIE and NARVAL, and we added new data acquired with the spectrograph MUSICOS\textsuperscript{\footnotemark}. We applied the Least-Squares Deconvolution technique to obtain Stokes V and Stokes I mean profiles to compute longitudinal magnetic field for a subset. Moreover, for the same subset, we analysed the Ca II H and K emission lines to calculate the S-index. We also derived atmospheric parameters and Li abundances for all eighteen stars of our sample. We found that stars previously classified as RGB may actually be at a different evolutionary state. Furthermore, we identified that most stars in our sample with detection of surface magnetic field show at least moderate rotation velocities, but nonetheless, we could not detect a magnetic field in two fast rotators. Due to our small sample of magnetic giants, it is difficult to determine if the presence of surface magnetic field and the Li-rich giant phenomena could be somehow linked. The large variation of the radial velocity of part of our sample indicates that some of them might have a binary companion, which may change the way we look at the Li problem in giant stars.
\end{abstract}


\footnotetext{Based in part on observations made at the Pico dos Dias Observatory OPD/LNA.}



\section{Introduction}

From the standard models of stellar evolution, low-mass stars should begin their lives with lithium (Li) abundance close to the meteoritical value ($\sim3.2$ dex) and then deplete this initial amount throughout its life without getting any enhancement. By the time a low-mass star ($\lesssim3.0\,M_{\odot}$) reaches the Red Giant Branch (RGB), processes such as the deepening of the convective zone, together with the event known as First Dredge-Up (hereafter FDU), should decrease drastically the amount of Li abundance remaining in the star, by a factor of about 60 for stars with solar metallicity \citep{iben_a_1967,iben_b_1967}. At this point of the evolution, $A(Li)\leq1.5$ dex is expected in standard models. This range of abundances is observed in $\sim99$\% of the red giant stars studied in several different surveys (e.g., \citealt{brown1989,pilachowski2000,kumar2011,charbonnel2020}). Nevertheless, the $\sim1$\% of red giants that present a high Li abundance are the ones that attract the most attention and interest of the astronomical community. The reason is that, to this day, researchers debate about the source of the Li enhancement, that can be an internal production or external contamination. Furthermore, within each of these scenarios, the understanding of the physical processes occurring in the star, which can, in fact, explain the anomalous abundances we observe, is also an unsettled issue (e.g., \citealt{charbonnel2020}).

On the side of external sources of Li enrichment, engulfment of planets and brown dwarfs are often indicated as a process capable of explaining the emergence of Li-rich giant stars \citep{siess1999,aguilera_gomez2016}. However, this form of Li enrichment can only account for abundances up to $\sim2.2$ dex \citep{aguilera_gomez2016}, which leaves without explanation a large number of Li-rich giants. Additionally, Li-enrichment by planet engulfment can only occur in the RGB climb \citep{casey2019}, making it difficult to justify the existence of Li-rich Red Clump giants. Therefore, internal synthesis emerges as the most likely explanation for stars with Li abundances greater than $2.2$ dex, or even more evident, super Li-rich stars with $A(Li)\geq3.2$ dex. It is worth highlighting that tidal spin-up of a binary companion is an external source capable of causing internal disturbances in the stellar mixing process. This issue is discussed by \cite{casey2019}.

The most reliable mechanism to explain how new Li can be produced in a red giant star was proposed by \cite{cameron1971}. Nonetheless, researchers do not seem to agree about the necessary conditions or events that would trigger a non-standard mixing capable of making the \citeauthor{cameron1971} mechanism actually work to generate anomalous Li abundances in the stellar photospheres. Among those conditions and events we can highlight: signs of magnetic activity and detection of surface magnetic fields \citep{fekel1993,guandalini2009,lebre2009,auriere2015}; the presence of a stellar companion generating tidal effects \citep{denissenkov2004,casey2019}; and engulfment of a substellar companion causing the increase of the star's rotation velocity \citep{siess1999,drake2002,denissenkov2004}. Moreover, disruptive stages such as the luminosity function RGB bump \citep{charbonnel2000} and the helium (He) flash \citep{kumar2011,Kumar2020} could create by itself the necessary physical conditions for a non-canonical mixing in the star's convective zone, but there is no clear understanding of what happens inside the star in those critical moments.

From the first discovery of a Li-rich red giant \citep{wallerstein1982} until recently, no more than 200 Li-rich giants had been catalogued \citep{casey2016}. However, studies analysing hundreds of thousands of stars from different surveys are revealing a much larger number of this type of objects \citep{deepak2019,casey2019}, which may lead to breakthroughs in what we know until now about the subject, especially for stars with seismological data \citep{singh2019}. Studying in close detail low-mass stars in the Red Clump phase, \cite{Kumar2020} found that those objects present a much higher Li abundance than what is predicted by stellar evolution models, which indicates that the Li enrichment phenomenon in giant stars must be much more common than previously thought. Based on that discovery, \cite{Kumar2020} propose that a much lower Li abundance should be used to classify upper-RGB and Red Clump stars as Li rich. Finally, the Gaia Data Release 2 (hereafter Gaia DR2, \citealt{gaia_collaboration2018}) have also created a singular opportunity for the reanalysis of the evolutionary status of several known Li-rich giants, and the understanding of how these stars fit in this new scenario containing thousands of Li-rich giant stars is of fundamental importance.

In this study we used 18 targets that we had access to spectra, all Li-rich giants, coming from \cite{charbonnel2000,lebre2009,kumar2011}. We reevaluate the H-R diagram for 17 Li-rich giant stars by using the new parallaxes and magnitudes from the Gaia DR2. One additional star studied by \cite{lebre2009}, HD 218153, was added to our analysis, even though it has a normal Li content. As we explain in Section 2, this star serves as a control target regarding the other magnetic Li-rich giants. We compute the atmospheric parameters, Li abundances and radial velocity (RV) of those stars by using the spectral analysis tool iSpec \citep{blanco_cuaresma2014,blanco_cuaresma2019}. We also apply Least-Squares Deconvolution (LSD) technique \citep{donati1997, kochukhov2010} to generate high signal-to-noise LSD profiles in order to evaluate the magnitude of the longitudinal magnetic field ($B_{l}$) in a subset with spectropolarimetric data available at PolarBase \citep{petit2014}. In addition, we compute the S\textsubscript{index} for the same subset. Furthermore, we study the RV of the stars in our sample to check if its variation can signal the presence of a binary companion. We divide the next sections as follows: in Section 2 we describe how we selected our sample, and the instruments utilised in the observations. In Section 3 we present the methods used to study our sample. In Section 4 we show the results obtained in our study, which are discussed in Section 5. Finally, we draw some conclusions in Section 6.

\section{Sample selection and observational data}

\begin{table}
	\centering
    \caption{Parallaxes and magnitudes of the stars in our sample from Gaia DR2 \citep{gaia_collaboration2018}. The distance (d) in Parsec is given by $10^{3}/\pi_{G}$. $G$ is the stellar magnitude in the Gaia band and $A_{G}$ is the line-of-sight extinction in the $G$-band.}
    \label{tab:01}
    \begin{tabular}{c c c c c c c}
    	\hline
    	Star ID & $\pi_{G}$ & d & $G$ & $A_{G}$ \\
    	 & (mas) & (pc) & (mag) & (mag) \\
    	\hline
   		HD 9746 & $6.346\pm0.091$ & $157.6^{+2.3}_{-2.2}$ & 5.456 & 0.368 \\[1.5ex]
        HD 21018 & $2.097\pm0.094$ & $476.9^{+22.4}_{-20.5}$ & 6.166 & --- \\[1.5ex]
        HD 30834 & $6.334\pm0.269$ & $157.9^{+7.0}_{-6.4}$ & 4.207 & 0.472 \\[1.5ex]
        HD 31993 & $2.727\pm0.041$ & $366.7^{+5.6}_{-5.4}$ & 7.077 & 0.398 \\[1.5ex]
        HD 33798 & $2.135\pm1.037$ & $468.4^{+442.4}_{-153.1}$ & 6.668 & --- \\[1.5ex]
        HD 39853 & $4.818\pm0.185$ & $207.6^{+8.3}_{-7.7}$ & 4.937 & 0.608 \\[1.5ex]
        HD 63798 & $5.262\pm0.036$ & $190.0^{+1.3}_{-1.3}$ & 6.234 & --- \\[1.5ex]
        HD 90633 & $8.823\pm0.024$ & $113.3^{+0.3}_{-0.3}$ & 6.002 & 0.105 \\[1.5ex]
        HD 112127 & $7.794\pm0.036$ & $128.3^{+0.6}_{-0.6}$ & 6.534 & 0.321 \\[1.5ex]
        HD 116292 & $11.097\pm0.105$ & $90.1^{+0.9}_{-0.8}$ & 5.050 & 0.367 \\[1.5ex]
        HD 126868 & $26.881\pm0.182$ & $37.2^{+0.3}_{-0.3}$ & 4.596 & --- \\[1.5ex]
        HD 170527 & $5.701\pm0.023$ & $175.4^{+0.7}_{-0.7}$ & 6.721 & 0.134 \\[1.5ex]
        HD 205349 & $1.341\pm0.070$ & $745.7^{+41.1}_{-37.0}$ & 5.528 & 0.823 \\[1.5ex]
        HD 214995 & $11.699\pm0.076$ & $85.5^{+0.6}_{-0.6}$ & 5.583 & 0.333 \\[1.5ex]
        HD 217352 & $4.526\pm0.059$ & $220.9^{+2.9}_{-2.8}$ & 6.760& 0.247 \\[1.5ex]
        HD 218153 & $4.199\pm0.038$ & $238.2^{+2.2}_{-2.1}$ & 7.304 & 0.144 \\[1.5ex]
        HD 232862 & --- & --- & 9.694 & --- \\[1.5ex]
        HD 233517 & $1.138\pm0.057$ & $878.7^{+46.3}_{-41.9}$ & 9.266 & 0.229 \\
    	\hline
    \end{tabular}
\end{table}

A total of 18 stars of spectral type K and G were selected to compose our sample. Those stars were studied by different works in the last two decades including \cite{charbonnel2000,lebre2009,kumar2011}. The Gaia parameters for our sample stars are shown in Table \ref{tab:01} and their atmospheric parameters found in the literature are shown in Table \ref{tab:09}. The giant star HD 218153 was studied by \cite{lebre2009} and we have decided to include it in our sample, even though it is not Li-rich. However, this star is a fast rotator and present surface magnetic field. Those attributes make HD 218153 a good target to contrast with some other stars of our sample that are also fast rotators and have surface magnetic field detected but show anomalously high Li abundance in their photospheres. As it is shown in Table \ref{tab:01}, HD 232862 is the only star that does not have any parallax measurement and, as a consequence, we could not derive its luminosity.

\subsection{Instruments}

We gathered spectra of known lithium-rich giant stars in the archives of two spectropolarimeters, ESPaDOnS and NARVAL, and two spectrographs, ELODIE \citep{moultaka2004} and SOPHIE, in order to compose our sample. Moreover, we made observations for three stars of our sample with the spectrograph MUSICOS (Table \ref{tab:02}).

From the 18 targets studied in this work, we used spectra of either ESPaDOnS or NARVAL for 9 of them (Table \ref{tab:02}). The twin echelle spectropolarimeters ESPaDOnS and NARVAL are mounted, respectively, at the CFHT (Canada-France-Hawaii Telescope), in Hawaii (US), and at the Bernard Lyot Telescope, on the Pic du Midi (France). Despite the fact that ESPaDOnS was built to be assembled at a 3.6 m telescope, and NARVAL to be assembled at a 2.0 m telescope, they have almost the exact same instrumental specifications. They can operate at three different observation modes with distinct resolving power (R): star-only (R = 76 000), star+sky (R = 65 000), polarimetric (R = 65 000). The first two modes record only the intensity spectrum. On the other hand, the polarimetric mode can measure the circular and linear polarisation as well.

In order to complete our sample, we have collected spectra of Li-rich giants in the archives of the echelle spectrographs ELODIE and SOPHIE. The ELODIE spectrograph stayed active from June 1993 to August 2006 at the 1.93 m telescope of Haute Provence Observatory, in France. This instrument was replaced by the spectrograph SOPHIE, which is still functional. Some of the improvements that SOPHIE present in comparison with ELODIE are the higher resolving power and better radial velocity (RV) precision. While ELODIE could achieve a spectral resolution of 42 000, SOPHIE can achieve a resolution of 75 000 when is on high spectral resolution mode. The other mode of SOPHIE, with spectral resolution of 40 000, assure higher throughput when observing fainter objects. On this study, we analysed spectra from both ELODIE and SOPHIE spectrographs.

Finally, we observed the targets HD 21018, HD 214995 and HD 217352 with the echelle spectrograph MUSICOS. This instrument is assembled at the 1.60 m Perkin-Elmer telescope located at the Pico dos Dias Observatory (OPD, Brazil). The spectrograph MUSICOS has a resolving power of 35,000 and a spectral coverage between 3800 and 8800 \AA\ (approximately 100 orders) in two expositions: the ``blue" exposition goes from 3800 \AA\ to 5400 \AA\ and the ``red" exposition goes from 5400 \AA\ to 8800 \AA. Our observations were made only in the red part of the spectral coverage.

\begin{table}
	\centering
    \caption{Atmospheric parameters and Li abundance found in the literature for the stars in our sample.}
    \begin{threeparttable}
    \label{tab:09}
    \begin{tabular}{c c c c c c c}
    	\hline
    	Star ID & $T_{\text{eff}}$ & log g & [Fe/H] & A(Li) &  Ref.\tnote{b}\\
    	 & (K) & (dex) & (dex) & (dex) & \\
    	\hline
   		HD 9746   & 4490 & 2.14 & -0.10 & 3.73\tnote{a} & TT17 \\
        HD 21018  & 5327 & 2.05 &  0.07 & 2.93\tnote{a} & TT17 \\
        HD 30834  & 4283 & 1.79 & -0.24 & 2.63\tnote{a} & TT17 \\
        HD 31993  & 4500 & ---  &  0.10 & 1.40 & CB00  \\
        HD 33798  & 4500 & ---  & -0.30 & 1.50 & CB00 \\
        HD 39853  & 3900 & 1.16 & -0.30 & 2.80 & KM11 \\
        HD 63798  & 5000 & 2.50 & -0.10 & 1.86 & KM11 \\
        HD 90633  & 4600 & 2.30 &  0.02 & 1.98 & KM11 \\
        HD 112127 & 4340 & 2.10 &  0.09 & 3.01 & KM11  \\
        HD 116292 & 5050 & 3.00 & -0.01 & 1.50 & KM11 \\
        HD 126868 & 5500 & ---  & -0.07 & 2.40 & CB00 \\
        HD 170527 & 4842 & 2.57 & -0.35 & 3.24\tnote{a} & TT17 \\
        HD 205349 & 4138 & 0.89 & -0.18 & 1.88\tnote{a} & TT17 \\
        HD 214995 & 4626 & 2.43 &  0.04 & 3.06\tnote{a} & TT17 \\
        HD 217352 & 4570 & 2.53 &  ---  & 2.64 & KM11 \\
        HD 218153 & 5000 & 3.00 & -0.15 & 0.00 & LB09 \\
        HD 232862 & 4938 & 3.79 & -0.20 & 1.88\tnote{a} & TT17 \\
        HD 233517 & 4475 & 2.25 & -0.37 & 4.11 & KM11 \\
    	\hline
    \end{tabular}
    \begin{tablenotes}
        \item[a] Abundances submitted to NLTE correction
        \item[b] References. CB00: \cite{charbonnel2000}; LB09: \cite{lebre2009}; KM11: \cite{kumar2011}; TT17: \cite{takeda2017}.
    \end{tablenotes}
    \end{threeparttable}
\end{table}

\begin{table*}
	\centering
    \caption{Atmospheric parameters and Li abundance computed in this work for the stars in our sample.}
    \begin{threeparttable}
    \label{tab:02}
    \begin{tabular}{c c c c c c c c}
		\hline
        Star ID & Instrument & $T_{\text{eff}}$ & log g & [Fe/H] & $\xi$ & A(Li)$_{\text{LTE}}$ & S/N\tnote{b} \\
         & & (K) & (dex) & (dex) & (km s$^{-1}$) & (dex) \\
        \hline
        HD 9746   & NARVAL   & $4605\pm30$  & $2.77\pm0.11$ & $-0.12\pm0.04$ & $1.90\pm0.06$ & $3.86\pm0.12$ & 569 \\
        HD 21018  & NARVAL   & $5675\pm135$ & $2.73\pm0.35$ & $0.14\pm0.09$  & $2.38\pm0.15$ & $3.41\pm0.24$ & 217 \\ 
        HD 30834  & ELODIE   & $4320\pm30$  & $1.87\pm0.22$ & $-0.31\pm0.05$ & $2.00\pm0.07$ & $2.62\pm0.18$ & 162 \\
        HD 31993  & NARVAL   & $4520\pm65$  & $2.78\pm0.20$ & $-0.10\pm0.08$ & $1.94\pm0.15$ & $1.57\pm0.22$ & 263 \\
        HD 33798  & NARVAL   & $4975\pm105$ & $3.51\pm0.14$ & $-0.10\pm0.09$ & $1.67\pm0.17$ & $1.67\pm0.26$ & 354 \\
        HD 39853  & ELODIE   & $3910\pm60$  & $1.59\pm0.20$ & $-0.50\pm0.08$ & $1.91\pm0.09$ & $2.51\pm0.34$ & 121 \\
        HD 63798  & ELODIE   & $5060\pm70$  & $2.83\pm0.23$ & $-0.14\pm0.07$ & $1.57\pm0.10$ & $1.94\pm0.14$ & 211 \\
        HD 90633  & ELODIE   & $4745\pm55$  & $2.80\pm0.19$ & $0.09\pm0.06$  & $1.61\pm0.08$ & $2.26\pm0.16$ & 173 \\
        HD 112127 & ELODIE   & $4645\pm55$  & $3.04\pm0.12$ & $0.37\pm0.06$  & $1.74\pm0.07$ & $3.67\pm0.19$ & 212 \\
        HD 116292 & ELODIE   & $5020\pm75$  & $3.02\pm0.14$ & $-0.03\pm0.07$ & $1.51\pm0.09$ & $1.65\pm0.17$ & 228 \\
        HD 126868 & ELODIE   & $5860\pm200$ & $3.98\pm0.27$ & $0.05\pm0.11$  & $1.57\pm0.22$ & $2.75\pm0.19$ & 215 \\
        HD 170527 & ELODIE   & $4805\pm105$ & $2.85\pm0.34$ & $-0.52\pm0.11$ & $1.83\pm0.22$ & $3.47\pm0.29$ & 166 \\
        HD 205349 & ELODIE   & $4285\pm25$  & $1.25\pm0.17$ & $-0.20\pm0.05$ & $3.30\pm0.06$ & $1.75\pm0.15$ & 129 \\
        HD 214995 & NARVAL   & $4795\pm30$  & $3.09\pm0.07$ & $0.07\pm0.03$  & $1.64\pm0.03$ & $3.46\pm0.13$ & 360 \\
        HD 217352 & NARVAL   & $4760\pm105$ & $3.11\pm0.20$ & $-0.12\pm0.10$ & $1.89\pm0.19$ & $2.86\pm0.32$ & 345 \\
        HD 218153 & NARVAL   & $4630\pm85$  & $2.64\pm0.22$ & $-0.35\pm0.08$ & $2.05\pm0.17$ & 0.30\tnote{a} & 174 \\
        HD 232862 & NARVAL   & $5085\pm110$ & $4.54\pm0.15$ & $-0.15\pm0.07$ & $1.88\pm0.28$ & $2.67\pm0.24$ & 154 \\
        HD 233517 & ESPaDOnS & $4485\pm60$  & $2.31\pm0.24$ & $-0.31\pm0.07$ & $1.98\pm0.13$ & $4.19\pm0.28$ & 125 \\
        \hline
        HD 21018  & MUSICOS & $5700\pm90$  & $2.97\pm0.18$ & $0.16\pm0.06$  & $2.02\pm0.09$ & $3.40\pm0.11$ & 248 \\
        HD 214995 & MUSICOS & $4785\pm55$  & $2.71\pm0.14$ & $-0.02\pm0.05$ & $1.29\pm0.07$ & $3.19\pm0.09$ &  85 \\
        HD 217352 & MUSICOS & $5025\pm200$ & $3.02\pm0.30$ & $-0.06\pm0.18$ & $1.42\pm0.25$ & $3.14\pm0.32$ &  44 \\
        \hline
	\end{tabular}
    \begin{tablenotes}
    	\item[a] Upper limit of the Li abundance.
    	\item[b] At 550 nm, approximately.
    \end{tablenotes}
    \end{threeparttable}
\end{table*}

\section{Methods}

\subsection{Deriving stellar luminosities from the Gaia parallaxes and magnitudes}

After obtaining effective temperatures ($T_\text{eff}$) from spectroscopy, we used the data and equations presented by \cite{andrae2018} to compute magnitudes for the stars in our sample. The stellar luminosities used in the H-R diagrams of Figures \ref{fig:03} and \ref{fig:04} were computed as follows. First we computed the bolometric correction (BC) (Equation \ref{eq:01}) of the magnitudes measured in the Gaia photospheric band (G-band). 

\begin{equation}
	BC_{G}(T_{\text{eff}})=\sum_{i=0}^{4}a_{i}\left(T_{\text{eff}}-T_{\text{eff}\odot}\right)^{i}
    \label{eq:01}
\end{equation}

The $a_{i}$ coefficients are reported by \cite{andrae2018}. It is worth highlighting that errors in $T_\text{eff}$ are not included in the determination of bolometric corrections.

From the Gaia parallaxes ($\pi_{G}$) we derived the distances in Parsec ($d=10^{3}/\pi_{G}$). The apparent magnitudes in the Gaia band ($G$), together with the line-of-sight extinction ($A_{G}$), are also provided. Therefore, we use Equation \ref{eq:10} to obtain the absolute magnitudes ($M_{G}$) for the stars in our sample.

\begin{equation}
    M_{G} = G - 5\log_{10}(d) + 5 - A_{G}
    \label{eq:10}
\end{equation}

The values of $d$, $G$ and $A_{G}$ used to derive $M_{G}$ are shown in Table \ref{tab:01}. Then, we use Equation \ref{eq:02} to calculate stellar luminosities.

\begin{equation}
	\log_{10}{\mathcal{L}}=-\frac{2}{5}\left(M_{\text{G}}+BC_{\text{G}}(T_{\text{eff}})-M_{\text{bol}\odot}\right)
    \label{eq:02}
\end{equation}

In our computation we set $M_{\text{bol}\odot}$ equal to $4.74$ mag. The calculated values for $\mathcal{L}$ are given in solar units. The data treatment that resulted in the $G$ fluxes of the Gaia Data Release 2, used in Equation \ref{eq:10}, can be seen in \cite{riello2018}. The error in luminosity comes exclusively from parallax errors. The errors in $T_{\text{eff}}$ shown in Figures \ref{fig:03} and \ref{fig:04} were estimated from our spectral synthesis, as we explained in Section 3.2.

\subsection{Atmospheric parameters and Li abundances}

We used the spectral analysis tool iSpec\footnote{For more information visit \url{https://www.blancocuaresma.com/s/iSpec}} to derive the atmospheric parameters and Li abundances for all stars in our sample. From the spectra available for each target, we always picked the spectrum with the highest S/N to compute the values shown in Table \ref{tab:02}. Furthermore, we added to the end of Table \ref{tab:02} the atmospheric parameters and Li abundances derived from our own observations of three targets with the spectrograph MUSICOS, as mentioned in Section 2.

In order to compute the atmospheric parameters, we made spectral synthesis in the optical region, from 4800 {\AA} to 6800 {\AA}, for all 18 stars of our sample. In this process, we also used a list of 279 atomic lines that is native from iSpec. The following atmospheric parameters are presented in Table \ref{tab:02}, along with their respective errors: effective temperature ($T_{\text{eff}}$), surface gravity (log g), metallicity ([Fe/H]), and microturbulence velocity ($\xi$).

\begin{figure}
    \centering
	\includegraphics[width=\columnwidth]{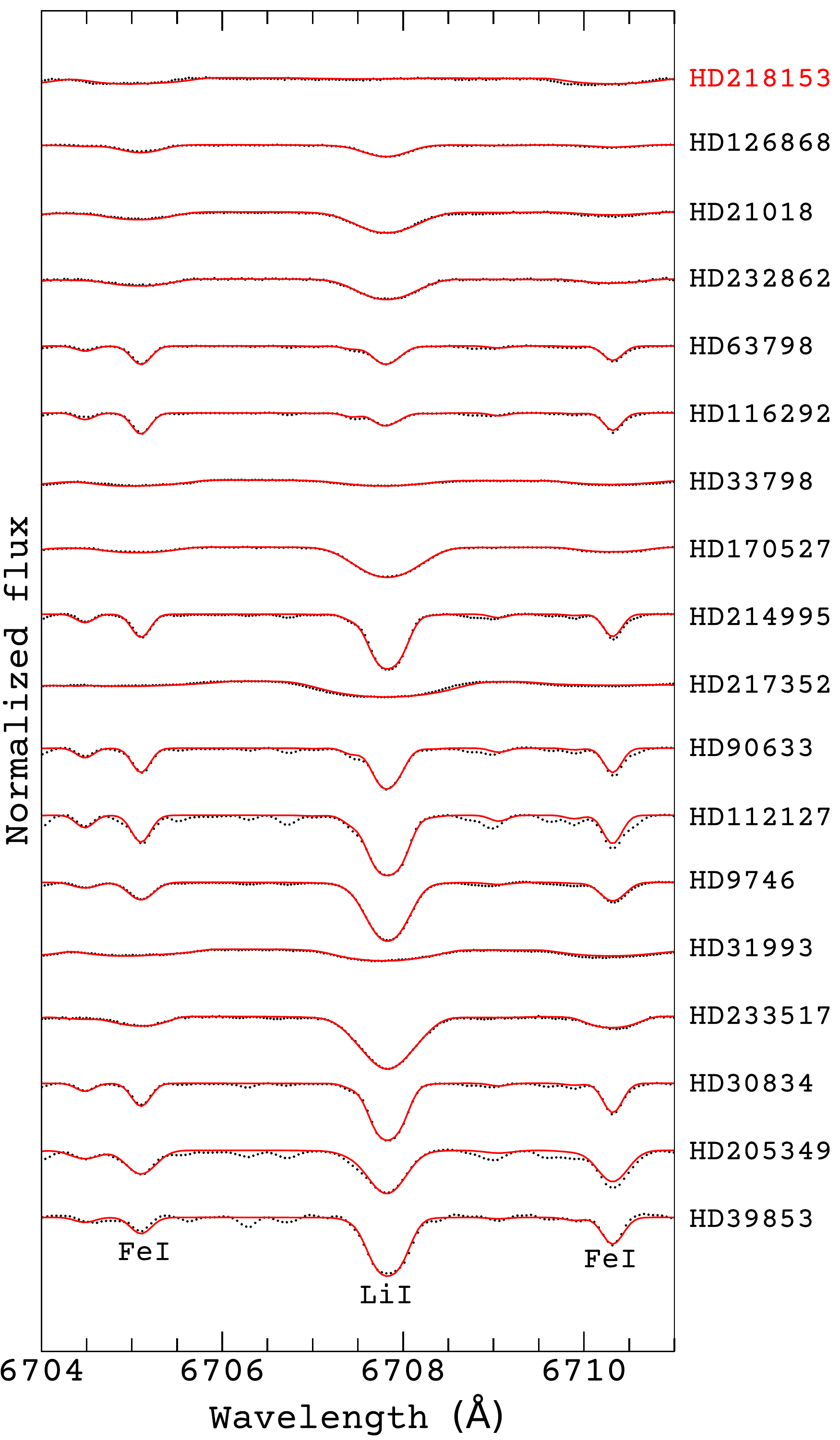}
    \caption{Spectral synthesis around the Li I resonance line at 6707.8 {\AA} for the 18 stars of our sample. Two Fe I lines are also identified in this plot. We assemble this panel to show the spectra of our sample stars increasing in $T_{\text{eff}}$ from the coolest (at the bottom) to hottest (at the top), except for the Li normal giant HD218153, which has been deliberately moved to the top of the plot.}
    \label{fig:01}
\end{figure}

On the other hand, for computing Li abundances, we used a much shorter spectral range, from 6702 {\AA} to 6712 {\AA}. Inside this interval, we were able to measure Li I abundances at the 6707.8 {\AA} resonance line. We also added five iron (Fe) lines to better adjust the spectral synthesis to the observed spectrum. In Figure \ref{fig:01} we show the spectral region used for computing Li abundances for the 18 stars of our sample, together with the identification of the 6707.8 {\AA} Li I line and the two most prominent Fe I lines. The spectral synthesis is represented by the red lines fitting the black dots, which form the observed spectrum. For HD 218153, we only measured an upper limit for the Li abundance and its spectrum is shown at the top of Figure \ref{fig:01}. 

For the spectral synthesis in both above mentioned cases, iSpec also gives to the user the choice of opting among five different radiative transfer codes, four different atomic line database, solar abundances from seven different authors and publications, and several different model atmospheres. We used MOOG \citep{sneden1973} as our radiative transfer code; Gaia-Eso Survey line list (version 5.0) as our source of atomic line data; Grevesse 2007 \citep{grevesse2007} as our source of solar abundances; and the MARCS Gaia-Eso Survey \citep{gustafsson2008} as our grid of model atmospheres.

\subsubsection{Carbon-nitrogen abundance ratio}

Molecular lines such as CN and CO are tiny and tricky to analyse. With few exceptions, most of the literature presents studies of carbon and nitrogen molecular lines in the near-infrared region (e.g., \citealt{hawkins2018}), which is a region not covered in our spectra. The ELODIE and SOPHIE spectra are restricted to the visible range, a region very poor in molecular lines of carbon and, especially, nitrogen. The NARVAL and ESPaDOnS spectra go a little into the near-infrared range ($\sim1000$ \AA), but it is still difficult to find nitrogen molecules lines strong enough to be studied. Despite that, we analysed our spectra using the spectroscopic tool iSpec in order to determine [C/H] and [N/H]. Using the VALD (Vienna Atomic Line Database), iSpec was capable of identifying several carbon lines (C2 I) and, eventually, a couple of nitrogen lines (N I). We then performed spectral synthesis in the regions where the lines were found, being able to determine [C/H] and [N/H] for 11 stars, two of which with high uncertainties. We then calculate the difference between [C/H] and [N/H] to derive [C/N]. The results are shown in Figure \ref{fig:05}.

\subsection{Radial Velocity}

We also focused on the radial velocity in order to check if the Li-rich giant stars in our sample have a normal degree of binary/multiple systems. We used the cross-match correlation technique \citep{pepe2002,allende_prieto2007} implemented inside the spectroscopic tool iSpec to derive the radial velocities of our target stars. The cross-match correlation algorithm is built around the Equation \ref{eq:04}.

\begin{equation}
    C\left(\upsilon\right) = \sum_{lines}\sum_{pix} p\left(pix,\upsilon\right)\cdot flux\left(pix\right)
    \label{eq:04}
\end{equation}

In this equation, $C\left(\upsilon\right)$ is the cross-correlation function, $p\left(pix,\upsilon\right)$ is a template function containing a line list of specific atomic/molecular transitions and $flux\left(pix\right)$ is the observed spectrum from which we compute the radial velocity. In our case, we have used the mask line list of Arcuturs Atlas, since we are dealing with giant stars.

After the velocity profile is built, the mean velocity is calculated by fitting a second order polynomial near the peak of the profile and additional parameters are calculated from a Gaussian fit \citep{allende_prieto2007}.

The errors in the radial velocity are calculated from the Equation \ref{eq:05}, based on the work of \cite{zucker2003}.

\begin{equation}
    \sigma_{\upsilon}^{2} = -\left[N\frac{C''\left(\upsilon\right)}{C\left(\upsilon\right)}\frac{C^{2}\left(\upsilon\right)}{1-C^{2}\left(\upsilon\right)}\right]^{-1}
    \label{eq:05}
\end{equation}

In this equation, $N$ is the number of bins in the spectrum, $C$ is the cross-correlation function and $C''$ is the second order derivative of the cross-correlation function. The RV computed for our sample are shown in the tables of Appendix \ref{app:01}.

\subsection{Longitudinal magnetic field measurements}

\begin{figure}
    \centering
    \includegraphics[width = \columnwidth]{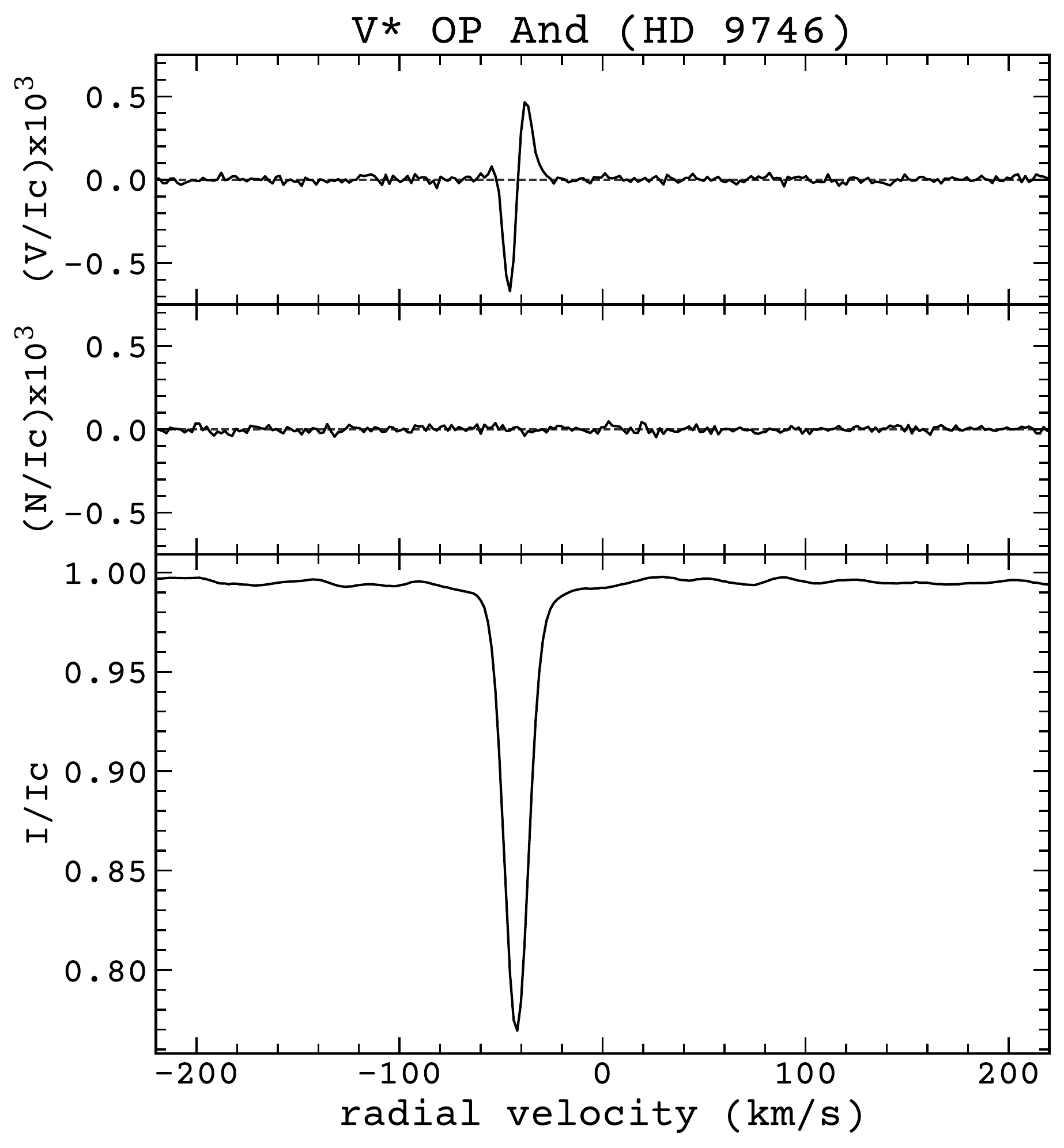}
    \caption{One of the LSD profiles analysed in this study. HD 9746 has a clear surface magnetic field detection characterised by the Zeeman signature seen in the LSD Stokes V mean profile at the top panel. In the middle panel is presented the null polarisation profile (N), and the bottom panel shows the LSD Stokes I mean profile.}
    \label{fig:02}
\end{figure}

In order to reconstruct Stokes I and V mean profiles with high enough S/N to be studied, we have applied the Least-Squares Deconvolution (LSD) technique to our spectropolarimetric data. This technique was first proposed by \cite{donati1997} and later revised by \cite{kochukhov2010}. In Figure \ref{fig:02} we show the result of applying the LSD technique to the spectropolarimetric data of HD 9746. It is possible to see the mean LSD Stokes V profile on the top panel, the mean LSD Stokes I profile on the bottom panel, and the diagnostic null profile (N), also known as the null polarisation profile, on the middle panel. 

Using mean profiles with enhanced S/N (e.g., the profile shown in Figure \ref{fig:02}) produced through the application of the LSD technique to the observed spectra, it is possible to compute the surface-averaged longitudinal magnetic field ($B_{l}$) using Equation \ref{eq:03}, presented by \cite{donati1997}.   

\begin{equation}
	B_{l} = - 2.14\cdot10^{11}\frac{\int\upsilon{V}(\upsilon)d\upsilon}{\lambda_{0}{g}_{0}{c}\int\left[1-{I}(\upsilon)\right]d\upsilon}
	\label{eq:03}
\end{equation}

In this equation, we have two integrals in the (radial) velocity space ($\upsilon$). The first integral, in the numerator, integrates the Stokes V LSD profile; and the second one, in the denominator, integrates the Stokes I LSD profile. The central wavelength $\lambda_{0}$ and the effective Land\'{e} factor $g_{0}$ represent arbitrary quantities used for the normalisation of the Stokes V LSD weights \citep{kochukhov2010}, that we have adjusted specifically for each LSD profile; and $c$ is the speed of light. We advise the reader to check the sections on the LSD technique and the computation of $B_{l}$ in the works by \cite{marsden2014} and \cite{auriere2015} for a more broadly coverage of this subject. The results of our determination of $B_{l}$ are shown in Appendix \ref{app:01} (Tables \ref{tab:03} to \ref{tab:07}).

\subsection{S-index}

The S-index is one of the most known and broadly used magnetic activity proxies. Its roots trace back to the 1960's, when the first long-term monitoring of chromospheric activity in dwarf stars took place at the Mount Wilson observatory \citep{wilson1978,duncan1991}, through the study of the Ca II H and K emission lines. Other indices that measure the level of the chromospheric activity in other regions of the spectrum have been conceived through the years, but the S-index kept its importance and relevance. Even though part of the spectra coming from NARVAL and ESPaDOnS used in this work were already normalised, as the Ca II H and K emission lines lie at the low efficient part of the spectrum, we proceed standard calibrations before we computed the S-index. The Doppler shift correction was performed using the radial velocities for each spectrum. We also performed a re-normalisation of the region of the H and K emission lines, as the continuum flux level is higher in the K order than the H order for some spectrum. As the S-index measurement depends on the flux of Ca II H and K lines to be normalised by the continuum flux in the nearby regions, this would have an effect on the measurements. This re-normalisation was performed by fitting a linear fit to the two separated continuum regions nearby the Ca II H and K lines and normalising the whole region with the best fit.

The S-index is a dimensionless quantity calculated by the ratio of fluxes of the Ca II H and K lines and the flux of the nearby continuum. The Ca II emission lines are measured using two triangle bandpasses with a FWHM of 1.09 \AA\ centred on the Ca II H and K line cores (3968.47 \AA\ and 3933.66 \AA, respectively). The fluxes of the nearby continuum are measured using two 20 \AA-wide rectangle bandpasses, named R and V bandpasses, centred at 4001.07 \AA\ and 3901.07 \AA, respectively. In order to calibrate our results to the original Mount Wilson scale, we use Equation \ref{eq:06}.

\begin{equation}
    S_{index} = \frac{aF_{H}+bF_{K}}{cF_{R}+dF_{V}}+e
    \label{eq:06}
\end{equation}

The coefficients used were adjusted for cool giants as described by \cite{auriere2015} and \cite{tsvetkova2017}. The errors were computed by using the standard error propagation from the observation spectrum.

We were not able to measure the S-index for all spectra because some of them presented large discontinuities in the Ca II H and K region, and thus were impossible to re-normalize. All S-index measurements are presented in Tables \ref{tab:03} to \ref{tab:07}.

\section{Results}

The major point of our results is the relocation of our sample on the H-R diagram. Where before some of those stars were thought to be ascending the RGB \citep{charbonnel2000}, we now claim that they are, in fact, well passed FDU and the RGB luminosity bump. This outcome is of fundamental importance to the discussion of the Li-rich giant evolutionary status. 

\subsection{H-R diagrams}

\begin{figure*}
    \centering
	\includegraphics[width=\textwidth]{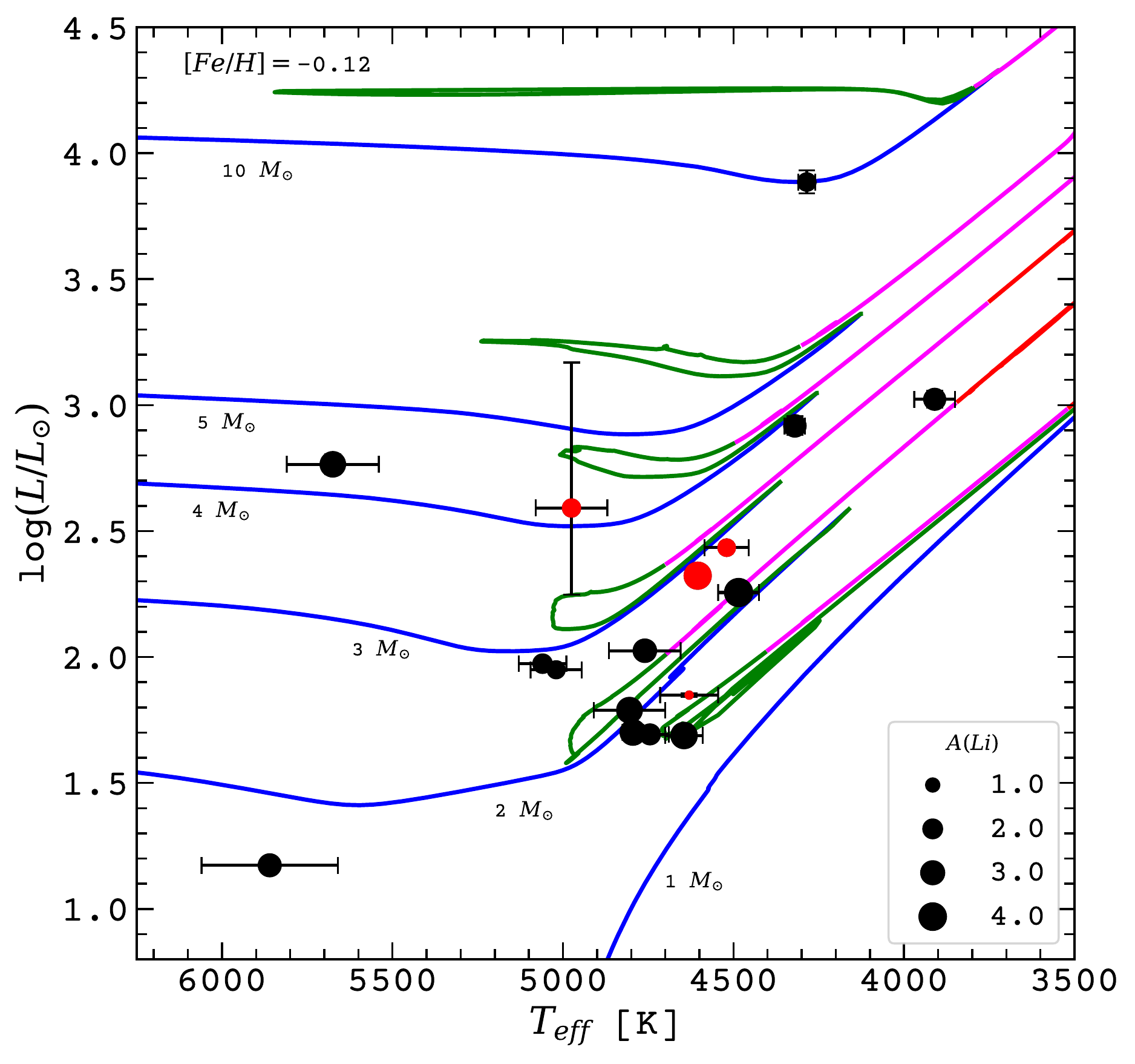}
	\caption{H-R diagram for our sample stars with Gaia parallax available. The star HD 232862 does not have parallax measurement. The red markers represent stars with definitive detection of surface magnetic field (DD). The size of each marker represents the Li content we computed for each star. We also present six different evolutionary tracks for six different masses, all of them computed with the median metallicity of our sample stars, [Fe/H] = $-0.12$ dex. The different colours in the evolutionary tracks are related to different stages of evolution: blue - Red Giant Branch (RGB); green - Red Clump or Horizontal Branch; pink - Early Asymptotic Giant Branch (E-AGB); red - Late Asymptotic Giant Branch (L-AGB).}
	\label{fig:03}
\end{figure*}

We present in this study H-R diagrams with evolutionary tracks that will help us to determine the most likely evolutionary state of our targets. In Figure \ref{fig:03} we present a general H-R diagram showing all the stars in our sample (except for HD 232862). The highlighted stars (red markers) are the ones that had magnetic field detected, and the symbol sizes represent the Li content for each star. Evolutionary tracks with different masses are plotted, all of which have [Fe/H] = $-0.12$ dex. This value is consistent with the median metallicity of our sample. All evolutionary tracks were taken from the MESA Isochrones and Stellar Tracks (version 1.2)\footnote{Available at \url{http://waps.cfa.harvard.edu/MIST/index.html}} \citep{choi2016}. 

\begin{figure*}
    \centering
    \includegraphics[scale=0.59]{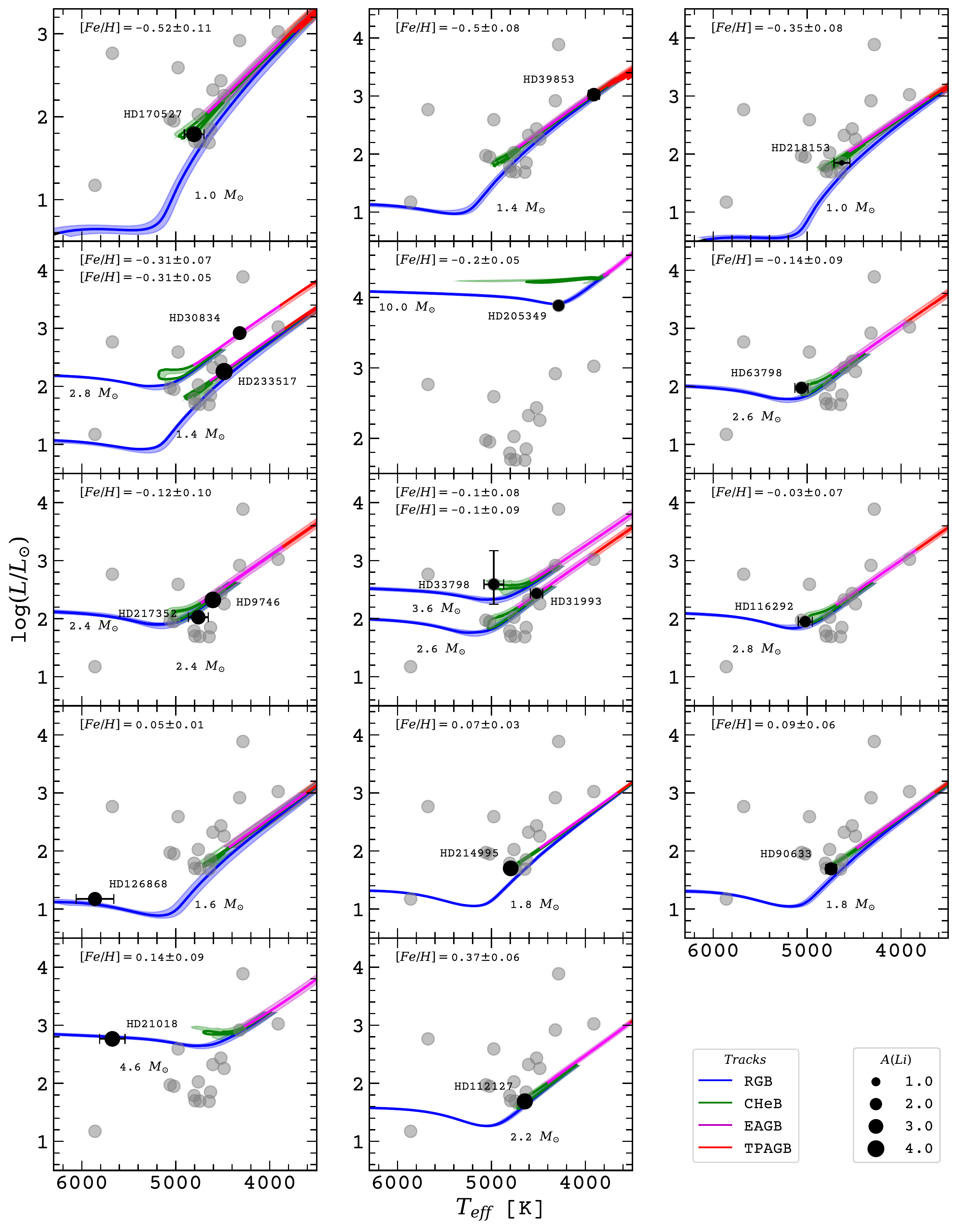}
    \caption{Individual H-R diagrams for each star of our sample and segregated by metallicities. The MESA evolutionary tracks were built specifically to the analysis of the evolutionary state of each star. We used the variable of mass to adjust the evolutionary tracks to the star's position on the H-R diagram. The metallicity considered for each track are shown in the panels. The shaded coloured regions indicate the errors in metallicity. In the last panel of the last row, we have two labels: one is indicating the Li content of our sample stars (same as Figure \ref{fig:03}), the other is the division of different evolutionary states (by colour) in the MESA evolutionary tracks. The acronyms mean the following: RGB - Red Giant Branch; CHeB - Core Helium Burning phase; EAGB - Early Asymptotic Giant Branch; TPAGB - Thermally Pulsating Asymptotic Giant Branch.}
    \label{fig:04}
\end{figure*}

While in Figure \ref{fig:03} we present a general view of our sample properties, in Figure \ref{fig:04} we present several H-R diagrams that are necessary to classify the evolutionary state of our sample stars. Each diagram of Figure \ref{fig:04} has an evolutionary track made specifically to the star's metallicity, and we have used the mass variable as a free parameter to adjust the evolutionary track to the marker position on the Luminosity-$T_{\text{eff}}$ plane. By doing this, we are also constraining the mass of the stars in our sample. All this care and specificity were necessary because it is quite difficult to differentiate RGB-bump from clump stars on the H-R diagram. In this way, we intend to make the most reliable classification possible based on evolutionary tracks. As explained in Section 5, we also make use of the $^{12}\text{C}/^{13}\text{C}$ and [C/N] abundance ratios in order to better determine if the stars have passed or not the FDU.

\subsection{Magnetic field and chromospheric activity}

Among the 18 stars of our sample, we investigate the presence of surface magnetic field for 7 of them: HD 9746, HD 21018, HD 31993, HD 33798, HD 218153, HD 232862, HD 233517, which are the stars we had access to spectropolarimetric measurements in PolarBase. Of those, we could make a Define Detection of surface magnetic field (DD) for 5 stars: HD 9746, HD 31993, HD 33798, HD 218153, HD 232862. Just four of the DD stars are represented in the H-R diagrams of Figures \ref{fig:03} and \ref{fig:04} (red makers), since HD 232862 does not have Gaia parallax yet. The results obtained from the spectropolarimetric analysis of the 5 stars with DD are shown in Appendix \ref{app:01} (Tables \ref{tab:03} to \ref{tab:07}), and the mean modulus of the longitudinal magnetic field ($|\overline{B_{l}}|$) is presented in Table \ref{tab:08}.

From the radial velocity (RV) analysis of the five DD stars, we cannot rule out the presence of a binary companion in any of them. However, the RV variation found in HD 9746 is most subtle and might be indicative of a less massive body in the system \citep{fischer2016}. Besides, the stars HD 218153 \citep{eker2008} and HD 232862 \citep{couteau1988,lebre2009} have already been mentioned in the literature as binary/multiple systems. Regarding the values computed for longitudinal magnetic field ($B_{l}$), we can observe a high variation of the magnitude of $B_{l}$ as well as in polarity (the sign of $B_{l}$), which indicates a dynamo nature of those magnetic fields \citep{lebre2009}.

From spectroscopy, we also measured the S-index for the stars we investigate the presence of surface magnetic field. The results for each observation day can be seen in Appendix \ref{app:01} (Tables \ref{tab:03} to \ref{tab:07}), and the mean values of the S-index are in Table \ref{tab:08}. The mean values obtained for the stars HD 21018 and HD 233517 come from observations carried out on the same night. As expected, the stars with DD presented a much higher S-index than the stars we could not detect any sign of surface magnetic field (nd).

\section{Discussion}

\subsection{Evolutionary status of the stars in our sample}

In this section, we investigate the evolutionary status of our sample. We consider in this analysis the H-R diagrams shown in Figure \ref{fig:04}. The MESA evolutionary tracks were built individually for each star, where the metallicities are the ones shown in Table \ref{tab:02} and the masses of the stars could be estimated by adjusting the tracks to the marker position on the Luminosity-$T_{\text{eff}}$ plane.

\begin{figure}
    \centering
    \includegraphics[width=\columnwidth]{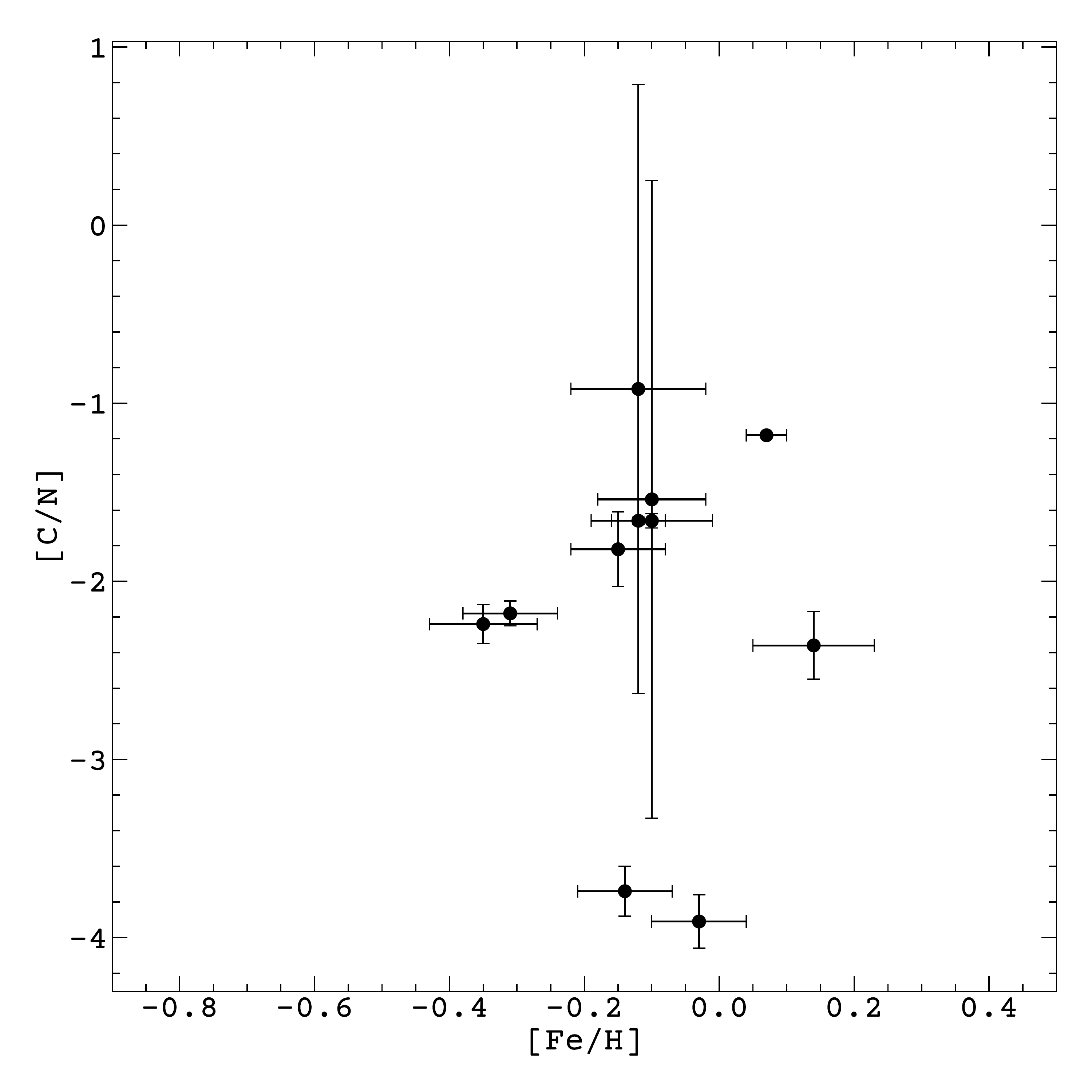}
    \caption{We present [C/N] as a function of [Fe/H] for 11 stars in our sample. Relaying on the evolutionary tracks, we classify two of our sample stars as RGB, HD 9746 and HD 217352. Exactly those two stars present the highest values of [C/N] (Table \ref{tab:08}). However, these abundances are still low for what is expected for a RGB star.}
    \label{fig:05}
\end{figure}

The main problematic about Li-rich giants is to define in which evolutionary state the enrichment of Li occurs, considering it happens indeed in a singular event. What is known from previous studies is that Li-rich giants tend to be agglomerated at the RGB bump, Red Clump and the early-AGB \citep{charbonnel2000,kumar2011,deepak2019}. Based on the position on the H-R diagram, obtained from the new Gaia parallaxes and photometry, we can determine in what of those regions our sample stars stand. We use the $^{12}\text{C}/^{13}\text{C}$ isotopic ratio available in the literature (Table \ref{tab:08}) in order to double-check if our targets have already undergone FDU or not. We also used the [C/N] abundance ratio that we computed for some of our sample stars (Table \ref{tab:08} and Figure \ref{fig:05}). In that manner, we can test if the evolutionary status we are obtaining agree with those stellar evolution spectroscopic constraints. 

\begin{table*}
    \centering
    \caption{Stellar parameters computed in this work and collected from other works that are relevant to the analysis of the results presented in Section 4 and to the discussions in Section 5. Besides the stellar luminosities, we show data regarding the carbon isotopic ratio ($^{12}\text{C}/^{13}\text{C}$), the carbon-nitrogen abundance ratio ([C/N]), the rotation velocity ($Vsin\,i$), the mean S\textsubscript{index} and the mean modulus of the longitudinal magnetic field ($|\overline{B_{l}}|$) for our sample stars. The column \textit{Det.} informs about the stars with define detection (DD) or no detection (nd) of surface magnetic field. Moreover, the column named \textit{Binarity} informs if the star is classified as binary system or not according to SIMBAD Astronomical Database.}
    \begin{threeparttable}
    \label{tab:08}
    \begin{tabular}{c c c c c c c c c}
        \hline
        Star ID & $\log(L/L_{\odot})$ & $^{12}\text{C}/^{13}\text{C}$\tnote{a} & [C/N]& $Vsin\,i$\tnote{b} & Det. & $\overline{\text{S}}_{\text{index}}$ & $|\overline{B_{l}}|$ & Binarity \\
         & (dex) & (dex) & (dex) & (km s$^{-1}$) & & & (G) & \\ 
        \hline
         HD 9746   & $2.323^{+0.013}_{-0.012}$ & 24  & $-1.66\pm0.02$ & 7.2  & DD  & $0.805\pm0.007$ & $5.9\pm0.1$  & No \\[1.5ex]
         HD 21018  & $2.765^{+0.040}_{-0.038}$ & --- & $-2.36\pm0.19$ & 21.2 & nd  & $0.262\pm0.006$ & ---  & Yes \\[1.5ex]
         HD 30834  & $2.918^{+0.038}_{-0.036}$ & 13  & --- & 2.3  & --- & ---   & ---  & No  \\[1.5ex]
         HD 31993  & $2.434^{+0.013}_{-0.013}$ & --- & $-1.54\pm1.79$ & 32.0 & DD  & $0.908\pm0.002$ & $9.7\pm1.2$ & No  \\[1.5ex]
         HD 33798  & $2.592^{+0.578}_{-0.344}$ & --- & $-1.66\pm0.04$ & 29.0 & DD  & $0.687\pm0.003$ & $8.8\pm0.6$  & No  \\[1.5ex]
         HD 39853  & $3.023^{+0.034}_{-0.033}$ & 6   & --- & 3.1  & --- & ---   & ---  & No  \\[1.5ex]
         HD 63798  & $1.974^{+0.006}_{-0.006}$ & 8   & $-3.74\pm0.14$ & ---  & --- & ---   & ---  & No  \\[1.5ex]
         HD 90633  & $1.693^{+0.003}_{-0.002}$ & 7   & --- & ---  & --- & ---   & ---  & No  \\[1.5ex]
         HD 112127 & $1.688^{+0.004}_{-0.004}$ & 19  & --- & 1.6  & --- & ---   & ---  & No  \\[1.5ex]
         HD 116292 & $1.949^{+0.008}_{-0.008}$ & --- & $-3.91\pm0.15$ & 3.7  & --- & ---   & ---  & No  \\[1.5ex]
         HD 126868 & $1.173^{+0.006}_{-0.006}$ & --- & --- & 14.4 & --- & ---   & ---  & No  \\[1.5ex]
         HD 170527 & $1.789^{+0.004}_{-0.004}$ & --- & --- & 22.9 & --- & ---   & ---  & No  \\[1.5ex]
         HD 205349 & $3.886^{+0.047}_{-0.044}$ & 9   & --- & 6.5  & --- & ---   & ---  & No  \\[1.5ex]
         HD 214995 & $1.701^{+0.006}_{-0.005}$ & 13  & $-1.18\pm0.01$ & 4.9  & --- & ---   & ---  & No  \\[1.5ex]
         HD 217352 & $2.025^{+0.011}_{-0.011}$ & 35  & $-0.92\pm1.71$ & 42.0 & --- & ---   & ---  & No  \\[1.5ex]
         HD 218153 & $1.849^{+0.008}_{-0.007}$ & --- & $-2.24\pm0.11$ & 29.4 & DD  & $0.924\pm0.038$ & $8.0\pm1.4$  & No  \\[1.5ex]
         HD 232862 & ---  & --- & $-1.82\pm0.21$ & 20.2 & DD  & $1.094\pm0.046$ & $36.0\pm1.5$ & Yes \\[1.5ex]
         HD 233517 & $2.256^{+0.045}_{-0.042}$ & 9   & $-2.18\pm0.07$ & 17.6 & nd  & $0.393\pm0.029$ & ---  & No  \\[1.5ex]
        \hline
    \end{tabular}
    \begin{tablenotes}
        \item[a] All values extracted from \cite{kumar2011} except for HD 217352 and HD 233517, obtained from \cite{charbonnel2000} and \cite{strassmeier2015}, respectively.
        \item[b] From \cite{takeda2017}: HD 9746; HD 21018; HD 30834; HD 170527; HD 205349; HD 214995; HD 232862. From \cite{charbonnel2000}: HD 33798; HD 39853; HD 112127; HD 126868; HD 233517. From \cite{kovari2013}: HD 217352. From \cite{kovari2016}: HD 218153. From \cite{kovari2017}: HD 31993. From \cite{rebull2015}: HD 116292.
    \end{tablenotes}
    \end{threeparttable}
\end{table*}

According to the numbers mentioned by \cite{gilroy1991}, stars not yet ascending the RGB have the $^{12}\text{C}/^{13}\text{C}>40$, which reveals an unmixed isotopic ratio. Meanwhile, stars that have already experienced mixing during the first dredge-up should present a lower $^{12}\text{C}/^{13}\text{C}$ ratio, something in the range of 20 - 25. However, a significant number of stars present a $^{12}\text{C}/^{13}\text{C}$ ratio lower than 20, which is not predicted by standard evolutionary models. This anomalous isotopic ratio abundance is often attributed to extra-mixing that might occur after the dredge-up \citep{charbonnel1998}. This scenario allows great amounts of the isotope $^{13}\text{C}$ to be carried to the star's convective zone, which would justify this large ratio decrease.

Relying on the evolutionary tracks of Figure \ref{fig:04}, we classify the stars of our sample in the following way: two subgiant stars, HD 21018 and HD 126868; two RGB stars, HD 9746 and HD 217352; eight Red Clump (CHeB) stars, HD 33798, HD 63798, HD 90633, HD 112127, HD 116292, HD 170527, HD 214995 and HD 218153; three early-AGB (EAGB) stars, HD 30834, HD 31993 and HD 233517; one late-AGB (TPAGB) star, HD 39853; one supergiant star, HD 205349.

This classification based on the evolutionary tracks is in agreement with the $^{12}\text{C}/^{13}\text{C}$ shown in Table \ref{tab:08}. For example, the stars HD 9746 and HD 217352 are classified as RGB stars but present different values of $^{12}\text{C}/^{13}\text{C}$ (24 dex and 35 dex, respectively). This may indicate that HD 9746 has already passed FDU, while HD 217352 has not. On the other hand, the [C/N] computed for those stars are very low and could indicate that they are also past the RGB tip. Meanwhile, the stars classified as Red Clump have a $^{12}\text{C}/^{13}\text{C}$ below 20 dex, and stars classified as being on the AGB have even lower $^{12}\text{C}/^{13}\text{C}$.

Some stars classified as Red Clump, due to either high parameters error bars or their position on the H-R diagram, could be classified into a different evolutionary state than the one given. Two examples that stand out are HD 33798 and HD 112127. Even though the former has a huge luminosity error and the latter is positioned between RGB and CHeB tracks, our approach is to classify them as Red Clump stars due to their low values of [C/N] ($-1.66\pm0.04$ dex) and $^{12}\text{C}/^{13}\text{C}$ (19 dex), respectively. Those cases similar to HD 112127 --- where error in $T_{\text{eff}}$ and [Fe/H] makes it difficult to classify stars as either RGB or Red Clump --- happen less severely for other stars, making it easier to determine their evolutionary states.

In \cite{charbonnel2000}, HD 9746, HD 112127 and HD 233517 are classified as RGB stars located at the luminosity bump and HD 31993, HD 33798 and HD 116292 are classified as RGB stars located before the luminosity bump. These results were obtained by using the old Hipparcos catalogue \citep{perryman1997} measurements. In our analysis, which use the Gaia DR2 parallaxes and photometry \citep{gaia_collaboration2018}, we have found that the stars HD 33798, HD 112127 and HD 116292 belong to the Red Clump; HD 31993 and HD 233517 are early-AGB stars; and HD 9746 is in fact ascending the RGB.

Finally, we have to point out that intermediate-mass stars ($3\,M_{\odot}\lesssim{M}\lesssim8\,M_{\odot}$) evolve in a slightly different way when compared to low-mass stars, and this issue can complicate the analysis that has been made until this point. That is because more massive stars can start He burning before the point where core matter becomes degenerate. Therefore, intermediate-mass stars may not go through extra-mixing while ascending the RGB, what can favour the survival of Li remnant from the main sequence \citep{charbonnel2000}; and it is also known that massive stars can synthesise Li through a process known as Hot Bottom Burning in the E-AGB \citep{lattanzio1996}. Nonetheless, the tipping point where those changes begin to happen is not associated to a sharp stellar mass value, but rather to a range of masses starting at $\sim2.2\,M_{\odot}$ and going up to $\sim4.0\,M_{\odot}$.

\subsection{Li-rich giant problem and binary systems}

It is no surprise that giant stars can present anomalous Li abundances, high magnetic activity, or huge rotation velocities. All of those features have been observed and reported in previous works \citep{fekel1993,drake2002,guandalini2009}. However, it is yet to come the time when we will be able to draw a clear picture of what is happening with these particular kind of stars that do not follow the standard theory of evolution. One characteristic very poorly explored in the past, but that has gained evidences recently \citep{casey2019} is whether the Li-rich stars have a normal rate of binarity, a question raised before by \cite{kumar2011}.

We have found that some stars in our sample show a degree of RV variation through the observation days. For some, we have enough measurements to question whether the system is single or not, while for others more observations are needed. We tend to associate low variation in RV to the presence of a planetary mass body and high variation in RV to the presence of a stellar mass body \citep{plavchan2015,fischer2016}. However, only a more detailed analysis can define the type of bodies present in the systems studied in this work (e.g., \citealt{massarotti2008}), which is beyond the scope of this paper, and should be the subject of a future work.   

The stars that show a very subtle change in RV are HD 9746, HD 112127 and HD 214995. The targets HD 21018, HD 30834, HD 39853, HD 63798, HD 90633, HD 116292, HD 126868, HD 170527, HD 205349 and HD 233517, have at most two measurements of RV, which turns any analysis of RV variation impracticable. Nevertheless, there is literature reporting HD 21018 as a binary system \citep{eker2008}. The remaining five stars (HD 31993, HD 33798, HD 218153, HD 232862 and HD 217352) show a significant change in RV throughout the observations, which makes us question if our sample of Li-rich giants is biased towards binary/multiple systems. The targets HD 31993, HD 33798, HD 218153 and HD 217352 present more than one peak in the velocity profile from the analysis of some spectra, which is indicative of a double-lined spectroscopic binary (SB2).

In this line of reasoning, a fair portion of our sample needs further investigation regarding its radial velocity variation and its relation with the Li problem. The high rotation velocities, the presence of magnetic fields, together with the anomalous Li abundances, could be much better understood, from a theoretical standpoint, if we introduce elements such as tidal spin up of a binary companion. We think this is the right path to go on in order to solve the puzzle. Nonetheless, further investigation has to be done. The data regarding radial velocity (Appendix \ref{app:01}), connected with models, can produce parameters such as orbital period and mass ratio. In this way, we will be able to characterise and confirm the binarity in some of our systems \citep{fulton2018}.

\section{Conclusions}

We have presented a new study of the evolutionary state for 18 stars under the perspective of the Gaia DR2, in opposition to the previous data from the Hipparcos catalogue \citep{perryman1997,van_leeuwen2007}. Moreover, we have also investigated the presence of surface magnetic field in 7 of them. Due to a RV variation of several km s$^{-1}$, we have also noticed that a significant portion of our targets may be composed by binary/multiple systems. Some of these targets even presented a double peak in the velocity profile, an issue that opens a window for further investigation of the Li-enrichment nature of those systems.

Our results have shown that Li-rich stars that were previously thought to be RGB stars \citep{charbonnel2000} are here identified with different evolutionary states, specially with the Red Clump. We have used MESA Isochrones and Stellar Tracks in order to determine the position of our stars on the H-R diagram. Furthermore, we have also used carbon-nitrogen atomic and isotipic ratios, particularly $^{12}\text{C}/^{13}\text{C}$ and [C/N], to confirm that our sample stars have passed the FDU phase and the RGB tip. The very low values of [C/N] computed for 11 stars in our sample indicate that all of them must have evolved past the RGB. However, we need to confirm these values using more reliable C and N lines, especially available in spectra that cover larger portions of the near-infrared region. With only the exception of HD 217352 ($^{12}\text{C}/^{13}\text{C}=35$), all the other stars have $^{12}\text{C}/^{13}\text{C}<25$, which confirms that FDU has happened. Nevertheless, HD 217352 could be a Red Clump star if it had only experienced a partial dredge-up while ascending the RGB, which is not unusual for stars with a mass between $2.2$ and $3.0\,M_{\odot}$, depending on the metallicity \citep{boothroyd1999}. Besides the intention to determine carbon isotopic abundances for all stars in our sample and also to better determine the $^{12}\text{C}/^{13}\text{C}$ ratios shown in Table \ref{tab:08}, we have also to emphasise the importance of improving the determination of other fragile elements such as $^7$Be \citep{carlberg2018}. That is important to identify with accuracy where external contamination of Li takes place, as well as to better constrain the conditions of Li depletion and production inside the stars.

Regarding the magnetism of our sample stars, it is not possible to identify a specific evolutionary state where the internal dynamo of those stars could be reignited, especially in such a small sample. However, we highlight that six out of seven stars of which we have investigated the presence of magnetic field present a moderate/high rotation velocity, a characteristic that could be linked to tidal spin up caused by a binary companion. We emphasise that the phenomenon of Li-enrichment is also possibly linked to the tidal spin-up of a binary companion. There are some other works in the literature presenting results that favour even further the scenario just mentioned, where the presence of a binary companion could increase the rotation velocity of the star and perhaps reactivate the internal dynamo of the stars. For example, even though we could not detect any sign of surface magnetic field, \cite{strassmeier2015} hypothesise that HD 233517 must have only developed a low-intensity surface magnetic field. Meanwhile, the models worked on in \cite{privitera2016} show that HD 9746 must actually have a high rotation velocity, about 30 $km\,s^{-1}$, while we are reporting a $Vsin\;i$ of only 7.2 $km\,s^{-1}$. All these issues add to the complexity to determine the nature of Li-rich giant stars, besides the limitation to differentiate systems with specific characteristics, which reinforces this long lasting problem in stellar astrophysics.

\section*{Acknowledgements}

The authors acknowledge the Brazilian research funding agencies CNPq and CAPES for financial support.

\section*{Data availability}

Part of the data underlying this article is in the public domain and available in: \textit{Gaia Archive} (\url{https://gea.esac.esa.int/archive/}), \textit{The ELODIE archive} (\url{http://atlas.obs-hp.fr/elodie/}), \textit{The SOPHIE archive} (\url{http://atlas.obs-hp.fr/sophie/}), \textit{PolarBase} (\url{http://polarbase.irap.omp.eu/}). Additional data underlying this article will be shared on reasonable request to the corresponding author.




\bibliographystyle{mnras}
\bibliography{example} 




\appendix

\section{Magnetic field, chromospheric activity and radial velocity data}
\label{app:01}

In this appendix we present in Tables \ref{tab:03} to \ref{tab:07} the results obtained for the stars in our sample analysed spectropolarimetrically and that have had definite detection (DD) of surface magnetic field. They are HD 9746, HD 31993, HD 33798, HD 218153, and HD 232862. Moreover, we also present the radial velocity for each observation day. The results are organised in the form of tables, where each column represent, from left to right, the following: Instrument; Universal Time Date of the observation; Heliocentric Julian Date of the observation; S/N of the LSD profiles; Number of mask lines used by LSD; Intensity of the longitudinal magnetic field ($B_{l}$); S$_{\text{index}}$; and Radial Velocity. The false alarm probability (FAP) is practically zero for all the LSD profiles analysed, which indicates Definite Detection (DD) of surface magnetic field. The specific atmospheric parameters of the masks used to compute the LSD profiles for each star with DD is shown in the caption of the tables. In Table \ref{tab:10} we present the radial velocity measurements for the rest of our sample.


\begin{table*}
    \centering
    \caption{Results of the spectropolarimetric data analysis of HD 9746. The atmospheric parameters of the mask ($\boldsymbol{M}$) are: $T_{\text{eff}}$ = 5000 K, log g = 3.00 dex, and solar metallicity. This star was observed by the spectropolarimeter NARVAL, located at the Telescope Bernard Lyot.}
    \label{tab:03}
    \begin{tabular}{c c c c c c c c}
         \hline
         Instrument & Date & HJD & S/N & No. of lines & $B_{l}$ & S$_{\text{index}}$ & RV \\
          & (UT) & 2 450 000 + & (LSD) & used & (G) & & (km s$^{-1}$) \\
         \hline 
         NARVAL & 16 Sep 2008 & 4 725.588 & 54 709 & 12 360 & $-14.7\pm0.4$ & --- & $-42.37\pm0.04$ \\
         NARVAL & 20 Sep 2008 & 4 730.462 & 31 280 & 12 362 &  $-4.6\pm0.7$ & --- & $-42.07\pm0.04$ \\
         NARVAL & 26 Sep 2008 & 4 735.537 & 41 341 & 12 359 &   $3.6\pm0.7$ & --- & $-42.11\pm0.04$ \\
         NARVAL & 28 Sep 2008 & 4 737.529 & 38 836 & 12 358 &   $6.7\pm0.6$ & $0.847\pm0.008$ & $-42.05\pm0.04$ \\
         NARVAL & 30 Sep 2008 & 4 739.551 & 50 321 & 12 359 &   $8.9\pm0.5$ & --- & $-42.24\pm0.04$ \\
         NARVAL & 21 Dec 2008 & 4 822.326 & 47 322 & 12 372 &   $6.1\pm0.5$ & $0.811\pm0.004$ & $-42.20\pm0.04$ \\
         NARVAL & 22 Jun 2010 & 5 369.616 & 33 454 & 12 364 &   $1.8\pm0.5$ & $0.763\pm0.008$ & $-42.11\pm0.04$ \\
         NARVAL & 23 Jun 2010 & 5 370.624 & 38 071 & 12 367 &   $0.8\pm0.3$ & $0.765\pm0.006$ & $-42.20\pm0.04$ \\
         NARVAL & 15 Jul 2010 & 5 392.631 & 43 936 & 12 373 &   $9.0\pm0.5$ & $0.757\pm0.004$ & $-42.05\pm0.04$ \\
         NARVAL & 24 Jul 2010 & 5 401.622 & 53 697 & 12 383 & $-12.2\pm0.4$ & $0.785\pm0.004$ & $-42.24\pm0.04$ \\
         NARVAL & 03 Aug 2010 & 5 411.639 & 51 402 & 12 372 &  $-6.3\pm0.3$ & $1.092\pm0.010$ & $-42.26\pm0.04$ \\
         NARVAL & 13 Aug 2010 & 5 421.644 & 41 474 & 12 373 &  $-1.2\pm0.7$ & $0.882\pm0.008$ & $-42.30\pm0.04$ \\
         NARVAL & 21 Aug 2010 & 5 429.608 & 39 745 & 12 371 &  $-0.8\pm0.5$ & $0.849\pm0.019$ & $-42.16\pm0.04$ \\ 
         NARVAL & 04 Sep 2010 & 5 443.681 & 52 662 & 12 370 &  $-0.3\pm0.4$ & $0.714\pm0.002$ & $-42.27\pm0.04$ \\
         NARVAL & 16 Sep 2010 & 5 455.631 & 19 130 & 12 382 &  $-3.3\pm1.1$ & $0.660\pm0.008$ & $-41.95\pm0.04$ \\
         NARVAL & 27 Sep 2010 & 5 466.524 & 39 122 & 12 370 & $-11.5\pm0.6$ & $0.780\pm0.004$ & $-42.25\pm0.04$ \\
         NARVAL & 13 Oct 2010 & 5 483.499 & 53 055 & 12 367 &  $-9.1\pm0.3$ & $0.762\pm0.003$ & $-42.45\pm0.04$ \\
         \hline
    \end{tabular}
\end{table*}

\begin{table*}
    \centering
    \caption{Results of the spectropolarimetric data analysis of HD 31993. The atmospheric parameters of the mask ($\boldsymbol{M}$) are: $T_{\text{eff}}$ = 5000 K, log g = 3.00 dex, and solar metallicity. This star was observed by the spectropolarimeter NARVAL, located at the Telescope Bernard Lyot.}
    \begin{threeparttable}
    \label{tab:04}
    \begin{tabular}{c c c c c c c c}
         \hline
         Instrument & Date & HJD & S/N & No. of lines & $B_{l}$ & S$_{\text{index}}$ & RV \\
          & (UT) & 2 450 000 + & (LSD) & used & (G) & & (km s$^{-1}$) \\
         \hline 
         NARVAL & 15 Sep 2008 & 4 724.637 & 16 578 & 12 353 &  $12.1\pm3.2$ & $0.874\pm0.002$ & $13.44\pm0.32$\tnote{*} \\
         NARVAL & 20 Sep 2008 & 4 729.710 & 15 999 & 12 365 &  $-6.8\pm3.4$ & $0.872\pm0.002$ & $15.04\pm0.35$ \\
         NARVAL & 26 Sep 2008 & 4 735.618 & 22 235 & 12 354 & $-14.1\pm2.4$ & $0.966\pm0.002$ & $16.91\pm0.34$ \\
         NARVAL & 29 Sep 2008 & 4 738.619 & 23 692 & 12 356 & $-11.2\pm2.3$ & $0.910\pm0.002$ & $19.16\pm0.48$ \\
         NARVAL & 01 Oct 2008 & 4 740.604 & 23 639 & 12 355 &  $-4.4\pm2.4$ & $0.920\pm0.002$ & $15.00\pm0.40$\tnote{*} \\
         \hline
    \end{tabular}
    \end{threeparttable}
\end{table*}

\begin{table*}
    \centering
    \caption{Results of the spectropolarimetric data analysis of HD 33798. The atmospheric parameters of the mask ($\boldsymbol{M}$) are: $T_{\text{eff}}$ = 5000 K, log g = 3.50 dex, and solar metallicity. This star was observed by the spectropolarimeter NARVAL, located at the Telescope Bernard Lyot.}
    \begin{threeparttable}
    \label{tab:05}
    \begin{tabular}{c c c c c c c c}
         \hline
         Instrument & Date & HJD & S/N & No. of lines & $B_{l}$ & S$_{\text{index}}$ & RV \\
          & (UT) & 2 450 000 + & (LSD) & used & (G) & & (km s$^{-1}$) \\
        \hline 
         NARVAL & 11 Mar 2007 & 4 171.299 & 28 018 & 11 957 & $-10.3\pm2.1$ & $0.713\pm0.002$ & $27.64\pm0.44$ \\
         NARVAL & 15 Mar 2007 & 4 175.321 & 21 687 & 11 957 & $-17.5\pm3.4$ & $0.766\pm0.006$ & $25.14\pm0.39$\tnote{*} \\
         NARVAL & 15 Set 2008 & 4 724.674 & 24 472 & 11 929 &  $-7.2\pm2.4$ & --- & $27.30\pm0.35$ \\
         NARVAL & 16 Sep 2008 & 4 725.679 & 32 748 & 11 937 & $-14.2\pm1.8$ & --- & $20.95\pm0.49$ \\
         NARVAL & 17 Sep 2008 & 4 726.648 & 31 625 & 11 924 & $-10.0\pm2.0$ & --- & $23.80\pm0.58$ \\
         NARVAL & 20 Sep 2008 & 4 729.676 & 28 824 & 11 927 &  $-1.5\pm2.4$ & --- & $22.88\pm0.66$ \\ 
         NARVAL & 21 Sep 2008 & 4 730.610 & 22 744 & 11 925 &  $-8.4\pm3.0$ & --- & $22.00\pm0.55$ \\
         NARVAL & 22 Sep 2008 & 4 731.606 & 29 369 & 11 926 &  $-8.6\pm2.6$ & --- & $25.15\pm0.81$ \\
         NARVAL & 25 Sep 2008 & 4 734.601 & 23 535 & 11 926 & $-13.2\pm2.9$ & $0.658\pm0.003$ & $27.07\pm0.48$ \\ 
         NARVAL & 26 Sep 2008 & 4 735.704 & 29 809 & 11 926 & $-15.3\pm2.4$ & --- & $22.73\pm0.72$ \\
         NARVAL & 27 Sep 2008 & 4 736.671 & 28 824 & 11 923 &  $-9.2\pm2.3$ & $0.635\pm0.003$ & $23.64\pm0.52$ \\
         NARVAL & 28 Sep 2008 & 4 737.613 & 26 879 & 11 924 &  $-7.2\pm2.4$ & $0.649\pm0.002$ & $29.43\pm0.39$ \\ 
         NARVAL & 29 Sep 2008 & 4 738.579 & 30 970 & 11 929 &  $-3.8\pm1.9$ & --- & $24.62\pm0.75$ \\
         NARVAL & 30 Sep 2008 & 4 739.685 & 32 522 & 11 937 &  $-4.1\pm1.8$ & --- & $21.49\pm0.31$ \\
         NARVAL & 01 Oct 2008 & 4 740.571 & 21 817 & 11 925 &  $-8.3\pm2.9$ & --- & $25.30\pm0.97$ \\ 
         NARVAL & 20 Dec 2008 & 4 821.349 & 25 524 & 11 935 &  $-2.6\pm2.3$ & $0.701\pm0.003$ & $25.00\pm0.47$ \\
         \hline
    \end{tabular}
    \begin{tablenotes}
        \item[*] Spectroscopic binary flag
    \end{tablenotes}
    \end{threeparttable}
\end{table*}

\begin{table*}
    \centering
    \caption{Results of the spectropolarimetric data analysis of HD 218153. The atmospheric parameters of the mask ($\boldsymbol{M}$) are: $T_{\text{eff}}$ = 5000 K, log g = 3.00 dex, and solar metallicity. This star was observed by the spectropolarimeter NARVAL, located at the Telescope Bernard Lyot.}
    \begin{threeparttable}
    \label{tab:06}
    \begin{tabular}{c c c c c c c c}
        \hline
         Inst. & Date & HJD & S/N & No. of lines & $B_{l}$ & S$_{\text{index}}$ & RV \\
          & (UT) & 2 450 000 + & (LSD) & used & (G) & & (km s$^{-1}$) \\
        \hline
         NARVAL & 05 Sep 2007 & 4 348.516 & 14 100 & 12 284 &  $3.1\pm5.1$ & $1.008\pm0.054$ & $-76.83\pm0.49$ \\
         NARVAL & 05 Sep 2007 & 4 349.501 & 15 634 & 12 284 &  $9.9\pm3.2$ & $0.963\pm0.040$ & $-77.73\pm0.40$ \\
         NARVAL & 14 Sep 2008 & 4 724.447 & 15 268 & 12 276 &  $7.9\pm3.0$ & --- & $-78.10\pm0.30$\tnote{*} \\
         NARVAL & 19 Sep 2008 & 4 729.508 &  9 485 & 12 268 &  $5.4\pm4.9$ & --- & $-74.65\pm0.38$ \\
         NARVAL & 25 Sep 2008 & 4 734.555 & 13 351 & 12 272 & $11.1\pm3.5$ & $0.931\pm0.054$ & $-82.33\pm0.55$\tnote{*} \\ 
         NARVAL & 26 Sep 2008 & 4 736.506 & 14 583 & 12 275 & $14.7\pm3.2$ & $0.849\pm0.039$ & $-81.56\pm0.41$ \\
         NARVAL & 29 Sep 2008 & 4 739.488 & 14 107 & 12 274 & $-3.6\pm3.3$ & $0.867\pm0.002$ & $-81.88\pm0.39$ \\
         \hline
    \end{tabular}
    \end{threeparttable}
\end{table*}

\begin{table*}
    \centering
    \caption{Results of the spectropolarimetric data analysis of HD 232862. The atmospheric parameters of the mask ($\boldsymbol{M}$) are: $T_{\text{eff}}$ = 5000 K, log g = 4.50 dex, and solar metallicity. This star was observed by both the spectropolarimeters NARVAL and ESPaDOnS, located, respectively, at the Telescope Bernard Lyot and at the Canada-France-Hawaii Telescope (CFHT).}
    \begin{threeparttable}
    \label{tab:07}
    \begin{tabular}{c c c c c c c c}
        \hline
         Instrument & Date & HJD & S/N & No. of lines & $B_{l}$ & S$_{\text{index}}$ & RV \\
          & (UT) & 2 450 000 + & (LSD) & used & (G) & & (km s$^{-1}$) \\
        \hline
         ESPaDOnS & 08 Dec 2006 & 4 077.937 & 12 628 & 11 260 & $-44.2\pm3.6$ & $1.142\pm0.019$ & $-0.19\pm0.32$ \\
         ESPaDOnS & 09 Dec 2006 & 4 078.751 & 10 223 & 11 260 & $-31.6\pm3.8$ & $1.144\pm0.025$ &  $1.21\pm0.23$ \\
         ESPaDOnS & 10 Dec 2006 & 4 079.826 &  6 092 & 11 260 & $-34.4\pm6.1$ & $1.221\pm0.060$ &  $0.76\pm0.38$ \\
         ESPaDOnS & 11 Dec 2006 & 4 080.817 & 11 549 & 11 260 & $-15.1\pm3.4$ & $1.163\pm0.021$ &  $0.81\pm0.47$ \\
         NARVAL   & 15 Sep 2008 & 4 724.574 &  6 995 & 10 715 & $-61.6\pm6.0$ & --- &  $0.31\pm0.22$ \\
         NARVAL   & 16 Sep 2008 & 4 725.634 &  9 908 & 10 942 & $-11.4\pm4.0$ & --- &  $1.15\pm0.29$ \\
         NARVAL   & 17 Sep 2008 & 4 726.607 &  8 235 & 10 940 & $-28.0\pm6.2$ & --- & $-0.72\pm0.28$ \\
         NARVAL   & 20 Sep 2008 & 4 729.636 &  8 900 & 10 940 & $-51.4\pm6.4$ & --- &  $0.00\pm0.27$ \\
         NARVAL   & 21 Sep 2008 & 4 730.548 &  5 710 & 10 652 & $-38.2\pm6.8$ & --- & $-0.36\pm0.42$ \\
         NARVAL   & 22 Sep 2008 & 4 731.566 &  8 915 & 10 938 & $-51.7\pm5.3$ & --- &  $0.27\pm0.29$ \\
         NARVAL   & 26 Sep 2008 & 4 735.578 &  5 877 & 10 607 & $11.1\pm10.9$ & --- &  $0.35\pm0.46$ \\
         NARVAL   & 27 Sep 2008 & 4 736.601 &  9 090 & 10 946 & $-45.7\pm6.0$ & $0.998\pm0.073$ & $-1.70\pm0.23$ \\
         NARVAL   & 28 Sep 2008 & 4 737.573 &  7 228 & 10 942 & $-32.0\pm5.4$ & $0.895\pm0.080$ & $-0.02\pm0.49$ \\ 
         NARVAL   & 29 Sep 2008 & 4 738.540 &  9 984 & 10 949 & $-48.8\pm5.1$ & --- & $-1.30\pm0.34$ \\
         NARVAL   & 30 Sep 2008 & 4 739.594 &  9 923 & 10 946 & $-42.0\pm5.7$ & --- &  $1.71\pm0.25$ \\
         NARVAL   & 30 Sep 2008 & 4 740.496 &  8 680 & 10 948 & $-28.4\pm6.3$ & --- &  $0.16\pm0.43$ \\
        \hline
    \end{tabular}
    \begin{tablenotes}
        \item[*] Spectroscopic binary flag
    \end{tablenotes}
    \end{threeparttable}
\end{table*}

\begin{table*}
    \centering
    \caption{Radial velocity measurements of the rest of our sample stars.}
    \begin{threeparttable}
    \label{tab:10}
    \begin{tabular}{c c c c c}
        \hline
        Star ID & Instrument & Date & JD & RV \\
         & & (UT) & 2 450 000 + & (km s$^{-1}$) \\
        \hline
        HD 21018  & NARVAL & 06 Sep 2007 & 4 349.620 & $10.06\pm0.18$ \\
        HD 21018 & MUSICOS & 21 Nov 2019 & 8 809.481 & $5.54\pm0.49$ \\
        \hline
        HD 30834  & ELODIE & 05 Oct 1996 & 361.608 & $-16.61\pm0.04$ \\
        \hline
        HD 39853  & ELODIE & 03 Oct 1996 & 359.647 & $82.02\pm0.04$ \\
        \hline
        HD 63798  & ELODIE & 14 Apr 2001 & 2 014.327 & $8.76\pm0.06$ \\
        HD 63798  & ELODIE & 21 Feb 2003 & 2 692.324 & $8.62\pm0.06$ \\
        \hline
        HD 90633  & ELODIE & 03 Nov 2003 & 2 946.686 & $-25.87\pm0.04$ \\
        \hline
        HD 112127 & ELODIE & 24 Apr 1997 & 563.383 & $6.43\pm0.03$ \\
        HD 112127 & ELODIE & 30 Jan 2002 & 2 304.619 & $6.53\pm0.03$ \\
        HD 112127 & ELODIE & 27 May 2003 & 2 787.418 & $6.69\pm0.03$ \\
        HD 112127 & SOPHIE & 26 Mar 2017 & 7 838.549 & $6.64\pm0.02$ \\
        \hline
        HD 116292 & ELODIE & 22 Feb 2003 & 2 692.618 & $-25.84\pm0.05$ \\
        \hline
        HD 126868 & ELODIE & 06 Jul 1998 & 1 001.360 & $-8.60\pm0.26$ \\
        \hline
        HD 170527 & ELODIE & 03 Nov 2003 & 2 947.256 & $-51.74\pm0.66$ \\
        \hline
        HD 205349 & ELODIE & 03 Oct 2003 & 2 916.304 & $-7.69\pm0.05$ \\
        HD 205349 & ELODIE & 15 Jan 2004 & 3 020.245 & $-7.51\pm0.05$ \\
        \hline
        HD 214995 & ELODIE & 10 Jul 2000 & 1 735.584 & $-28.34\pm0.05$ \\
        HD 214995 & SOPHIE & 27 Oct 2007 & 4 401.342 & $-28.28\pm0.04$ \\
        HD 214995 & NARVAL & 09 Sep 2007 & 4 353.487 & $-28.31\pm0.04$ \\
        HD 214995 & SOPHIE & 05 Nov 2015 & 7 332.288 & $-28.33\pm0.04$ \\
        HD 214995 & MUSICOS & 22 Nov 2019 & 8 809.544 & $-27.73\pm0.07$ \\
        \hline
        HD 217352 & NARVAL & 01 Oct 2012 & 6 202.427 & $-17.48\pm0.56$ \\
        HD 217352 & NARVAL & 02 Oct 2012 & 6 203.406 & $-14.57\pm0.58$\tnote{*} \\
        HD 217352 & NARVAL & 03 Oct 2012 & 6 204.429 & $-15.22\pm0.83$ \\
        HD 217352 & NARVAL & 04 Oct 2012 & 6 205.443 & $-13.39\pm0.76$ \\
        HD 217352 & NARVAL & 05 Oct 2012 & 6 206.431 & $-14.70\pm0.76$\tnote{*} \\
        HD 217352 & NARVAL & 12 Oct 2012 & 6 213.460 & $-16.29\pm0.39$\tnote{*} \\
        HD 217352 & NARVAL & 13 Oct 2012 & 6 214.335 & $-16.94\pm0.61$\tnote{*}\\
        HD 217352 & NARVAL & 23 Oct 2012 & 6 224.394 & $-17.96\pm0.43$\tnote{*} \\
        HD 217352 & NARVAL & 28 Oct 2012 & 6 229.349 & $-15.92\pm0.66$ \\
        HD 217352 & NARVAL & 29 Oct 2012 & 6 230.333 & $-13.91\pm1.44$ \\
        HD 217352 & NARVAL & 31 Oct 2012 & 6 232.347 & $-18.52\pm0.46$\tnote{*} \\
        HD 217352 & NARVAL & 02 Nov 2012 & 6 234.347 & $-14.23\pm0.43$\tnote{*} \\
        HD 217352 & NARVAL & 05 Nov 2012 & 6 237.405 & $-15.84\pm0.93$ \\
        HD 217352 & NARVAL & 06 Nov 2012 & 6 238.380 & $-19.19\pm0.82$ \\
        HD 217352 & NARVAL & 07 Nov 2012 & 6 239.276 & $-16.18\pm0.59$\tnote{*} \\
        HD 217352 & NARVAL & 19 Nov 2012 & 6 251.306 & $-17.49\pm0.50$ \\
        HD 217352 & NARVAL & 20 Nov 2012 & 6 252.296 & $-15.45\pm0.45$\tnote{*} \\
        HD 217352 & MUSICOS & 22 Nov 2019 & 8 809.557 & $-20.43\pm0.42$\tnote{*} \\
        \hline
        HD 233517 & ESPaDOnS & 09 Feb 2006 & 3 776.036 & $47.05\pm0.09$ \\
        HD 233517 & NARVAL   & 04 Apr 2008 & 4 561.470 & $46.70\pm0.13$ \\
        \hline
    \end{tabular}
    \begin{tablenotes}
        \item[*] Spectroscopic binary flag
    \end{tablenotes}
    \end{threeparttable}
\end{table*}


\bsp	
\label{lastpage}
\end{document}